\documentclass[conference]{IEEEtran}
\IEEEoverridecommandlockouts

\usepackage{cite}
\usepackage{graphicx}
\usepackage{url}
\usepackage{hyperref}
\usepackage{amsmath}
\usepackage{amssymb}
\usepackage{tikz}

\title{
Automated Post-Incident Policy Gap Analysis via Threat-Informed Evidence Mapping using Large Language Models
}

\author{
\IEEEauthorblockN{Oh Huan Lin, Jay Yong Jun Jie, Lee Ling Siu Mandy, Dr Jonathan Pan}
\IEEEauthorblockA{
Nanyang Technological University, Singapore \\
Emails: W240001@e.ntu.edu.sg, JAYY0002@e.ntu.edu.sg, MANDY001@e.ntu.edu.sg, 
JonathanPan@ntu.edu.sg 
}
}

\begin{document}
\maketitle

\begin{abstract}
Cybersecurity post-incident reviews are essential for identifying control failures and improving organisational resilience, yet they remain labour-intensive, time-consuming, and heavily reliant on expert judgment. This paper investigates whether Large Language Models (LLMs) can augment post-incident review workflows by autonomously analysing system evidence and identifying security policy gaps. We present a threat-informed, agentic framework that ingests log data, maps observed behaviours to the MITRE ATT\&CK framework, and evaluates organisational security policies for adequacy and compliance. Using a simulated brute-force attack scenario against a Windows OpenSSH service (MITRE ATT\&CK T1110), the system leverages GPT-4o for reasoning, LangGraph for multi-agent workflow orchestration, and LlamaIndex for traceable policy retrieval. Experimental results indicate that the LLM-based pipeline can interpret log-derived evidence, identify insufficient or missing policy controls, and generate actionable remediation recommendations with explicit evidence-to-policy traceability. Unlike prior work that treats log analysis and policy validation as isolated tasks, this study integrates both into a a unified end-to-end proof-of-concept post-incident review framework. The findings suggest that LLM-assisted analysis has the potential to improve the efficiency, consistency, and auditability of post-incident evaluations, while highlighting the continued need for human oversight in high-stakes cybersecurity decision-making.
\end{abstract}

\section{Introduction}
Cybersecurity post-incident reviews are essential for identifying control failures, understanding attacker behaviour, and improving organisational security posture \cite{Connolly2025PostIncident, AtlassianPostmortem}. In practice, these reviews remain largely manual and expert-driven, particularly when correlating large volumes of system logs with organisational security policies \cite{Antunes2022Audit, Hanson2021ICS}. As cyber threats evolve rapidly, static and infrequently reviewed policies increasingly struggle to reflect real-world attack techniques and operational conditions \cite{Linkov2014Risk, Hill2021Bruteforce}.

Recent advances in Large Language Models (LLMs) have demonstrated strong capabilities in reasoning over semi-structured data and orchestrating multi-step analytical tasks, suggesting potential applicability in post-incident analysis workflows \cite{Huang2023ReasoningSurvey, Wei2022CoT}. While prior work has explored LLMs for isolated tasks such as log analysis or policy compliance checking \cite{Zhang2025Survey, Cadet2024Compliance}, limited research has examined their use in integrated post-incident review workflows that connect technical evidence with governance-oriented policy evaluation. 

This paper investigates whether an LLM-driven, agentic workflow can support post-incident reviews by analysing log-derived evidence, mapping observed behaviours to the MITRE ATT\&CK framework, and identifying gaps in organisational security policies in a traceable and auditable manner. Rather than replacing human analysts, the approach is intended as a feasibility-focused, decision-support framework that augments evidence interpretation and policy evaluation.

The contributions of this paper are as follows:
\begin{itemize}
    \item An end-to-end, agentic post-incident review workflow integrating log analysis, threat attribution, and policy gap identification.
    \item A threat-informed evidence-to-policy mapping approach grounded in the MITRE ATT\&CK framework.
    \item An interpretable and auditable workflow design to support governance and compliance-oriented post-incident evaluation.
\end{itemize}

\section{Related Work}

\subsection{LLMs in Cybersecurity Analysis}
Recent studies demonstrate that Large Language Models (LLMs) can support a range of cybersecurity tasks, including log interpretation, vulnerability analysis, and partial audit automation \cite{Chin2025Audit}. Survey literature further reports that LLMs can enhance analyst productivity by reasoning over complex, semi-structured security data and coordinating multi-step analytical workflows \cite{Ruan2023Agents, Fang2024Hack}. Agentic LLM architectures have enabled systems that map vulnerabilities to the MITRE ATT\&CK framework or simulate attacker behaviour in controlled environments \cite{Jin2024Crimson}. 

While these approaches highlight the technical potential of LLMs, they primarily focus on isolated analytical tasks and do not explicitly connect observed incident evidence to organisational governance or security policy evaluation.

\subsection{AI-Assisted Audits and Policy Compliance}
An emerging body of work explores the use of LLMs to support security audits and compliance management \cite{Cadet2024Compliance, Salman2024Compliance}. Prior studies show that LLMs can assist in interpreting regulatory texts, mapping standards such as NIST or CIS controls to operational procedures, and automating portions of audit planning and reporting. Industry position papers similarly argue that LLM-based compliance tools can reduce manual effort and improve consistency in policy assessment \cite{Salman2024Compliance}. 

However, most existing approaches treat compliance as a document-centric task, focusing on policy-to-policy or control-to-standard comparisons. These systems rarely incorporate post-incident technical evidence, limiting their ability to assess whether documented controls remain effective under real attack conditions.

\subsection{Threat-Informed Post-Incident Auditing and Governance}
Traditional cybersecurity auditing frameworks adopt a top--down, policy-centric approach, where compliance is assessed against predefined controls through periodic and manually intensive reviews \cite{Antunes2022Audit}. Industry postmortem practices emphasize organisational learning and process improvement following incidents, but remain largely qualitative and human-driven, with limited mechanisms for evidence traceability or repeatable policy validation \cite{AtlassianPostmortem}. 

Threat modeling frameworks such as MITRE ATT\&CK provide a standardized abstraction layer between low-level system events and higher-level adversarial techniques \cite{MITREATTACK}. Despite this, existing audit and postmortem methodologies rarely operationalise threat-informed evidence mapping in an automated or auditable manner, leaving policy evaluation largely decoupled from observed attacker behaviour.

\subsection{Gaps in Automated Post-Incident Review}
Despite advances in LLM-based security analytics and compliance automation, fully integrated post-incident review workflows remain underexplored. Existing research typically addresses log analysis, threat detection, or policy compliance as separate problem domains, lacking end-to-end frameworks that unify evidence interpretation with policy gap identification. Practical challenges, including model hallucination, limited context windows, and insufficient traceability, further hinder adoption in audit-critical environments \cite{Loumachi2025RAG, Zhang2025Hallucination}. 

Traditional statistical and deep learning approaches to log analysis are effective for anomaly detection but lack the semantic reasoning required to relate observed behaviour to organisational policy adequacy \cite{Chourasiya2025Logs}. Few studies provide explicit mechanisms for linking analytical conclusions to verifiable log evidence and policy clauses, which is essential for forensic soundness and governance accountability. This work addresses this gap by integrating log-derived evidence, threat abstraction, and policy evaluation into a single, traceable LLM-assisted post-incident review framework.

\section{Research Aim and Scope}

The primary aim of this research is to evaluate whether Large Language Models (LLMs) can effectively augment cybersecurity post-incident review processes by autonomously analysing system evidence and identifying gaps in organisational security policies. Specifically, this study investigates the feasibility of using an agentic LLM-driven workflow to interpret log-derived evidence, map observed behaviours to established threat frameworks, and assess the adequacy of documented security controls. Rather than replacing human analysts, the research aims to determine whether LLMs can improve the efficiency, consistency, and traceability of post-incident evaluations by providing evidence-grounded and auditable insights that support cybersecurity governance and decision-making.

The scope of this study is limited to a simulated post-incident review scenario involving a brute-force authentication attack against a Windows OpenSSH service, mapped to the MITRE ATT\&CK technique T1110. The analysis focuses on Windows Event Log (EVTX) data as the primary source of technical evidence and evaluates organisational user account policies against baseline security controls. The system is implemented using a single large language model for reasoning and does not perform model fine-tuning or comparative benchmarking across multiple models. This research emphasises feasibility, interpretability, and workflow integration rather than detection accuracy or large-scale performance evaluation.

\section{Methodology}

\subsection{Experimental Design}
This study employs an experimental design to assess the feasibility of using a Large Language Model (LLM)-driven, agentic workflow to support cybersecurity post-incident reviews. The experiment simulates a realistic incident scenario in which system-generated evidence is analysed and compared against organisational security policies to identify control gaps. The workflow is structured to mirror the reasoning process of a human security analyst, progressing from evidence interpretation to threat attribution and policy evaluation. Evaluation focuses on interpretability, evidence traceability, and the ability to produce auditable and actionable outputs rather than detection accuracy or performance benchmarking.

\subsection{Agentic Workflow Architecture}
The proposed system is implemented as a multi-agent pipeline orchestrated using LangGraph. \cite{LangGraph2025, Taulli2025LangGraph} Each agent performs a distinct function, including log interpretation, threat attribution, policy retrieval, and policy gap identification, while maintaining a shared global state. GPT-4o is used as the primary reasoning model, and LlamaIndex supports semantic indexing and retrieval of organisational policy documents with line-level metadata. This modular architecture enables structured information flow across analysis stages and supports traceable reasoning throughout the post-incident review process.

Recent empirical evaluations of multi-step reasoning in large language models indicate that newer models exhibit greater resilience to reasoning complexity and reduced inconsistency across chained inference tasks, supporting their suitability for structured analytical workflows. \cite{Hoza2025Reasoning} Prior benchmarking studies in domain-specific classification tasks suggest that GPT-4o demonstrates strong consistency and reasoning performance, reinforcing its selection for evidence-driven reasoning in audit-critical contexts. \cite{Lin2025CancerLLM}

Unlike linear LLM pipelines, the agentic design allows intermediate outputs
to be validated and reused across stages, reducing redundant inference and
limiting error propagation. Persistent state management enables explicit
tracking of evidence, policy excerpts, and intermediate findings, which is
particularly important in audit and compliance contexts where traceability
is required.

\subsection{Incident Data and Threat Attribution}
The experiment uses Windows Event Log (EVTX) data representing a brute-force authentication attack against a Windows OpenSSH service. As EVTX files are stored in binary format, the logs are converted to XML and subsequently flattened into CSV format using a deterministic preprocessing script. The log analysis agent examines authentication-related events to identify temporal patterns and behavioural indicators of brute-force activity. Detected behaviours are mapped to the MITRE ATT\&CK framework, specifically Technique T1110 (Brute Force), providing a standardised threat context for policy evaluation. \cite{MITREATTACK}

Authentication logs were selected because they are commonly available in
enterprise environments and frequently used in incident investigations.
The selected dataset contains repeated authentication failures followed by
a successful login, a pattern commonly associated with credential guessing
attacks. This makes it suitable for evaluating both threat attribution and
policy adequacy, particularly for access control and authentication policies.

\subsection{Policy Evaluation and Evidence Traceability}
Organisational security policies and baseline best-practice controls are ingested in PDF format and indexed using LlamaIndex. Relevant policy clauses are retrieved based on semantic similarity to the detected threat behaviour. The policy evaluation agent compares the organisation’s controls against the baseline to identify insufficient or missing safeguards. Each identified gap is supported by explicit references to log-derived evidence and policy excerpts, along with a confidence assessment to support human validation. This design ensures that findings remain evidence-driven, auditable, and suitable for governance and compliance contexts in post-incident authentication policy evaluation.

Baseline security controls referenced in this study are derived from widely adopted governance frameworks, including NIST SP 800-53, ISO/IEC 27001, and the CIS Critical Security Controls, which define best-practice requirements for access control and credential management. \cite{NIST80053,ISO27001,CISControls}

\subsection{Implementation Details and Prompt Control}

To ensure reproducibility and audit suitability, the implementation places strict boundaries around data preprocessing, prompt structure, and model inference behaviour. All log preprocessing steps are deterministic and performed outside the Large Language Model (LLM). Windows Event Log (EVTX) files are converted to XML and subsequently flattened into CSV format using a fixed Python script, preserving key fields such as Event ID, timestamp, target account, and authentication status. This design avoids introducing variability at the data ingestion stage and ensures that identical evidence is supplied across repeated executions.

Prompting is structured at a high level to reflect the role of each agent in the workflow. Rather than using open-ended instructions, each agent receives task-specific prompts that constrain its responsibility, such as summarising authentication events, mapping observed behaviour to the MITRE ATT\&CK framework, or comparing retrieved policy clauses. Prompts explicitly instruct the model to reference observable evidence and retrieved policy text, discouraging speculative reasoning and unsupported conclusions.

Model inference is performed using a fixed temperature setting to minimise output variability across runs. Although minor variations in phrasing may still occur due to the probabilistic nature of LLMs, constraining sampling behaviour improves consistency in findings and supports repeatable analysis. This combination of deterministic preprocessing, structured prompting, and controlled inference contributes to the overall traceability and reliability of the proposed post-incident review workflow.

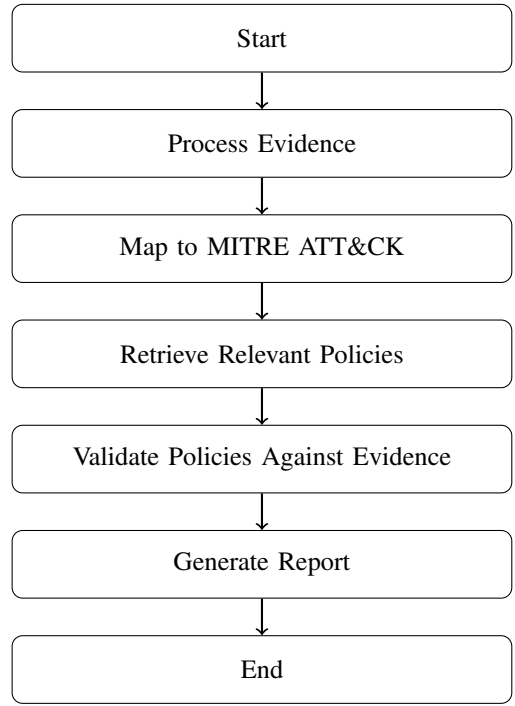
\begin{figure}[h]
\centering
\begin{tikzpicture}[
    node distance=1.4cm,
    every node/.style={
        draw,
        rounded corners,
        align=center,
        minimum width=0.75\linewidth,
        minimum height=0.9cm
    },
    arrow/.style={->, thick}
]

\node (start) {Start};

\node (evidence) [below of=start] {Process Evidence};

\node (mitre) [below of=evidence] {Map to MITRE ATT\&CK};

\node (policy) [below of=mitre] {Retrieve Relevant Policies};

\node (validate) [below of=policy] {Validate Policies Against Evidence};

\node (report) [below of=validate] {Generate Report};

\node (end) [below of=report] {End};

\draw[arrow] (start) -- (evidence);
\draw[arrow] (evidence) -- (mitre);
\draw[arrow] (mitre) -- (policy);
\draw[arrow] (policy) -- (validate);
\draw[arrow] (validate) -- (report);
\draw[arrow] (report) -- (end);

\end{tikzpicture}
\caption{End-to-end workflow for evidence-driven policy compliance analysis}
\end{figure}

\section{Findings}

\subsection{Threat Identification and Interpretation}
The LLM-driven workflow successfully identified the simulated brute-force authentication attack by analysing Windows Event Log patterns. Repeated failed logon attempts (Event ID 4625) followed by a successful authentication (Event ID 4624) within a short time window were consistently interpreted as indicative of brute-force behaviour. The model was able to summarise these patterns, with traceable reasoning, despite incomplete or noisy log fields, demonstrating an ability to reason over semi-structured evidence rather than relying on strict rule-based signatures. The identified behaviour was correctly mapped to MITRE ATT\&CK Technique T1110 (Brute Force), providing a standardised threat classification that supported downstream policy evaluation.

\subsection{Policy Gap Identification}
Based on the detected threat behaviour, the system retrieved relevant clauses from both the baseline policy and the organisation’s target policy and performed comparative analysis. The workflow identified multiple policy gaps, including an overly permissive account lockout threshold and an infrequent password rotation requirement. These findings were framed as risk-based deficiencies rather than binary compliance failures, highlighting how existing controls could be insufficient under observed attack conditions. Each gap was accompanied by a concise rationale and a recommended remediation aligned with industry best practices.

\subsection{Evidence-to-Policy Traceability}
All identified policy gaps were explicitly linked to supporting log evidence and policy excerpts. The system provided references to specific event patterns and policy clauses, enabling independent verification by human reviewers. This evidence-to-policy traceability reduced reliance on generative explanations and improved auditability. Confidence levels assigned to each finding reflected the quantity and consistency of supporting evidence, supporting prioritisation during review.

\subsection{Operational Implications}
The findings indicate that LLM-assisted, agentic workflows can meaningfully augment post-incident review processes by reducing manual effort while preserving interpretability and governance requirements. The structured outputs demonstrate potential for integration into audit and compliance workflows, particularly as a decision-support mechanism rather than a fully autonomous system. Human oversight remains essential, but the results suggest that LLMs can accelerate incident-driven policy evaluation and improve the consistency of post-incident analyses.

\section{Discussion}
The findings of this study demonstrate that Large Language Models (LLMs) can meaningfully augment cybersecurity post-incident reviews by bridging technical evidence analysis and organisational policy evaluation. Unlike traditional approaches that treat log analysis and compliance assessment as separate activities, the proposed agentic workflow integrates both within a single, traceable reasoning process. This integration enables incident-driven policy evaluation, allowing organisations to assess not only whether controls exist, but whether they are sufficient under observed attack conditions.

A key contribution of this work is the emphasis on evidence-to-policy traceability. By grounding each identified policy gap in verifiable log patterns and specific policy clauses, the system addresses a common barrier to adopting LLMs in audit-critical environments: trust. Rather than relying on opaque generative explanations, the workflow produces structured, auditable outputs that can be independently validated by human analysts. This positions LLMs as decision-support tools that enhance analyst efficiency while preserving accountability.

The results also highlight the practical role of agentic orchestration in managing complex reasoning tasks. By decomposing the post-incident review into modular stages, the system mirrors human analytical workflows and supports extensibility to additional evidence sources or policy domains. However, the study reinforces that LLM-assisted analysis should complement, not replace, human judgment. Expert oversight remains essential for contextual interpretation, risk acceptance decisions, and governance accountability. Overall, the findings suggest that LLM-driven post-incident review has strong potential to improve the timeliness, consistency, and governance alignment of cybersecurity operations when deployed within well-defined procedural and oversight frameworks.

\section{Limitations}
This study has several limitations that affect the generalisability and operational applicability of the proposed approach. First, the evaluation is limited to a single simulated incident scenario involving Windows Event Logs and a brute-force authentication attack. While this scenario is representative of common enterprise threats, the findings may not directly extend to more complex, multi-stage attacks or heterogeneous log sources such as network telemetry or cloud-native audit logs.

Second, the workflow relies on a single large language model for reasoning and does not include comparative benchmarking across alternative models or configurations. The probabilistic nature of LLM outputs introduces variability across runs, which may affect consistency in high-stakes audit contexts. Although structured prompts and validation steps mitigate this issue, full determinism is not guaranteed.

Third, the quality of policy gap identification is inherently dependent on the completeness and clarity of the organisational policy documents ingested. Ambiguous, outdated, or poorly structured policies may limit retrieval accuracy and lead to incomplete assessments. Additionally, the system does not perform quantitative performance evaluation or measure time savings relative to human analysts, focusing instead on feasibility and interpretability.

Finally, operational deployment raises broader considerations around data governance, privacy, and accountability. Incident logs and policy documents often contain sensitive information, requiring robust access controls and secure handling when integrated with external model APIs. These limitations indicate that while the approach is promising, further research is needed to validate scalability, reliability, and governance readiness in real-world enterprise environments.

\section{Future Work}
Future research can extend this work along several dimensions to improve robustness, scalability, and real-world applicability. First, the proposed workflow should be evaluated across a broader range of incident scenarios, including multi-stage attacks and heterogeneous evidence sources such as network logs, endpoint telemetry, and cloud audit trails. Incorporating cross-system correlation would enable more comprehensive incident reconstruction and strengthen policy evaluation in complex enterprise environments.

Second, future studies should explore controlled comparisons across multiple large language models and configurations to assess reasoning consistency, reliability, and cost-performance trade-offs. Techniques such as ensemble inference, deterministic decoding, or verification-based prompting could further reduce output variability in audit-critical contexts. Additionally, integrating retrieval-augmented generation more tightly into the reasoning process may improve grounding and reduce unsupported conclusions.

Third, quantitative evaluation metrics should be introduced to complement qualitative findings. These may include analyst time savings, consistency of findings across runs, and agreement between LLM-assisted outputs and expert assessments. Finally, future work should examine governance and deployment considerations, including secure on-premise or private-cloud deployments, access control mechanisms, and human-in-the-loop validation frameworks. Addressing these areas would support the transition of LLM-assisted post-incident review from proof-of-concept to operational use.

\section{Conclusion}
This paper investigated the feasibility of using Large Language Models (LLMs) to augment cybersecurity post-incident reviews through an evidence-driven, agentic workflow. By integrating log analysis, threat attribution, and policy evaluation within a single framework, the study demonstrates how LLMs can support incident-driven identification of security policy gaps with explicit evidence-to-policy traceability. The findings indicate that such workflows can improve the efficiency, consistency, and auditability of post-incident analysis while preserving the need for human oversight.

Rather than positioning LLMs as autonomous decision-makers, this work highlights their value as decision-support tools that assist analysts in navigating complex evidence and governance requirements. While limitations remain in scalability, determinism, and data governance, the results suggest meaningful potential for LLM-assisted post-incident review to enhance cybersecurity operations. This study lays a foundation for future research into trustworthy, transparent, and governance-aligned applications of agentic AI in security analysis.

\section{Code availability}
The GitHub repository will be made publicly available after acceptance of the paper at a peer-reviewed conference.

\bibliographystyle{IEEEtran}
\bibliography{references}

@article{Antunes2022Audit,
  author  = {Antunes, Miguel and Maximiano, Miguel and Gomes, Rui},
  title   = {A Client-Centered Information Security and Cybersecurity Auditing Framework},
  journal = {Applied Sciences},
  volume  = {12},
  number  = {9},
  pages   = {4102},
  year    = {2022},
  doi     = {10.3390/app12094102}
}

@misc{AtlassianPostmortem,
  author = {{Atlassian}},
  title  = {The Importance of an Incident Postmortem Process},
  year   = {n.d.},
  note   = {Online. Available: https://www.atlassian.com/incident-management/postmortem}
}

@article{Cadet2024Compliance,
  author  = {Cadet, Emmanuel and Etim, Etim D. and Essien, Isaac A. and others},
  title   = {Large Language Models for Cybersecurity Policy Compliance and Risk Mitigation},
  journal = {International Journal of Scientific Research in Humanities and Social Sciences},
  volume  = {1},
  number  = {2},
  pages   = {612--643},
  year    = {2024}
}

@misc{Chin2025Audit,
  author = {Chin, Jia Hao and Zhang, Peng and Cheong, Yong Xuan and Pan, Jonathan},
  title  = {Automating Security Audit Using Large Language Model Based Agent},
  year   = {2025},
  note   = {arXiv:2505.10732}
}

@article{Chourasiya2025Logs,
  author  = {Chourasiya, L. and Khatri, S. and Lilhore, U. K. and others},
  title   = {Advanced System Log Analyzer for Anomaly Detection and Cyber Forensic Investigations Using LSTM and Transformer Networks},
  journal = {Journal of Cloud Computing},
  volume  = {14},
  number  = {1},
  pages   = {60},
  year    = {2025},
  doi     = {10.1186/s13677-025-00789-y}
}

@misc{Connolly2025PostIncident,
  author = {Connolly, C.},
  title  = {Post-Incident Review: Boost Your Cybersecurity Resilience},
  year   = {2025},
  note   = {Cyber Defense Group. Available: https://www.cdg.io/blog/post-incident-review/}
}

@misc{Fang2024Hack,
  author = {Fang, Ruiqi and Bindu, R. and Gupta, A. and Zhan, Q. and Kang, D.},
  title  = {LLM Agents Can Autonomously Hack Websites},
  year   = {2024},
  note   = {arXiv:2402.06664}
}

@misc{Hanson2021ICS,
  author = {Hanson, Darin T.},
  title  = {Normalizing Cybersecurity: Improving Cyber Incident Response with the Incident Command System},
  year   = {2021},
  note   = {Homeland Security Digital Library}
}

@misc{Hill2021Bruteforce,
  author = {Hill, J.},
  title  = {New Research Shows Brute Force Attacks Rise 671\%},
  year   = {2021},
  note   = {Abnormal AI Blog}
}

@article{Hoza2025Reasoning,
  author  = {Hoza, Petr},
  title   = {Evaluating Reasoning in Large Language Models with a Modified Think-a-Number Game},
  journal = {Acta Informatica Pragensia},
  volume  = {14},
  number  = {2},
  pages   = {246--260},
  year    = {2025},
  doi     = {10.18267/j.aip.273}
}

@article{Huang2023ReasoningSurvey,
  author  = {Huang, Jie and Chang, Kai-Wei},
  title   = {Towards Reasoning in Large Language Models: A Survey},
  journal = {Findings of ACL},
  year    = {2023}
}

@article{Jin2024Crimson,
  author  = {Jin, Jun and Tang, Bo and Ma, Ming and others},
  title   = {Crimson: Empowering Strategic Reasoning in Cybersecurity through Large Language Models},
  journal = {ICCBD+AI},
  pages   = {18--24},
  year    = {2024},
  doi     = {10.1109/ICCBD-AI65562.2024.00011}
}

@misc{LangGraph2025,
  author = {{LangChain}},
  title  = {LangGraph: Build Resilient Language Agents as Graphs},
  year   = {2025},
  note   = {GitHub Repository}
}

@article{Lin2025CancerLLM,
  author  = {Lin, K.-H. and Kao, T.-H. and Wang, L.-C. and others},
  title   = {Benchmarking Large Language Models GPT-4o, Llama 3.1, and Qwen 2.5 for Cancer Genetic Variant Classification},
  journal = {NPJ Precision Oncology},
  volume  = {9},
  pages   = {141},
  year    = {2025},
  doi     = {10.1038/s41698-025-00935-4}
}

@article{Linkov2014Risk,
  author  = {Linkov, Igor and Anklam, Erika and Collier, Zachary A.},
  title   = {Risk-Based Standards: Integrating Top--Down and Bottom--Up Approaches},
  journal = {Environment Systems \& Decisions},
  volume  = {34},
  number  = {1},
  pages   = {134--137},
  year    = {2014},
  doi     = {10.1007/s10669-014-9488-3}
}

@article{Loumachi2025RAG,
  author  = {Loumachi, F. Y. and Ghanem, M. C. and Ferrag, M. A.},
  title   = {Advancing Cyber Incident Timeline Analysis Through Retrieval-Augmented Generation and Large Language Models},
  journal = {Computers},
  volume  = {14},
  number  = {2},
  pages   = {67},
  year    = {2025},
  doi     = {10.3390/computers14020067}
}

@misc{MITREATTACK,
  author = {{MITRE}},
  title  = {MITRE ATT\&CK Framework},
  year   = {2024},
  note   = {https://attack.mitre.org}
}

@misc{NIST80053,
  author = {{NIST}},
  title  = {Security and Privacy Controls for Information Systems and Organizations (SP 800-53)},
  year   = {2020}
}

@misc{ISO27001,
  author = {{ISO}},
  title  = {ISO/IEC 27001:2022 Information Security Management Systems},
  year   = {2022}
}

@misc{CISControls,
  author = {{Center for Internet Security}},
  title  = {CIS Critical Security Controls v8.1},
  year   = {2023}
}

@misc{Ruan2023Agents,
  author = {Ruan, Jiawei and Chen, Yi and Zhang, Bo and others},
  title  = {TPTU: Large Language Model-Based AI Agents for Task Planning and Tool Usage},
  year   = {2023},
  note   = {arXiv:2308.03427}
}

@article{Salman2024Compliance,
  author  = {Salman, A. and Creese, S. and Goldsmith, M.},
  title   = {Leveraging Large Language Models for Cybersecurity Compliance},
  journal = {IEEE European Symposium on Security and Privacy Workshops},
  pages   = {496--503},
  year    = {2024},
  doi     = {10.1109/EuroSPW61312.2024.00061}
}

@incollection{Taulli2025LangGraph,
  author    = {Taulli, Tom and Deshmukh, Girish},
  title     = {Introduction to LangGraph},
  booktitle = {Building Generative AI Agents},
  publisher = {Apress},
  year      = {2025}
}

@article{Wei2022CoT,
  author  = {Wei, Jason and Wang, Xuezhi and Schuurmans, Dale and others},
  title   = {Chain-of-Thought Prompting Elicits Reasoning in Large Language Models},
  journal = {arXiv},
  year    = {2022}
}

@article{Zhang2025Survey,
  author  = {Zhang, Jing and Bu, H. and Wen, H. and others},
  title   = {When LLMs Meet Cybersecurity: A Systematic Literature Review},
  journal = {Cybersecurity},
  volume  = {8},
  number  = {1},
  pages   = {55},
  year    = {2025},
  doi     = {10.1186/s42400-025-00361-w}
}

@article{Zhang2025Hallucination,
  author  = {Zhang, W. and Zhang, J.},
  title   = {Hallucination Mitigation for Retrieval-Augmented Large Language Models: A Review},
  journal = {Mathematics},
  volume  = {13},
  number  = {5},
  pages   = {856},
  year    = {2025},
  doi     = {10.3390/math13050856}
}

\end{document}